\documentclass[aps, superscriptaddress,nofootinbib]{revtex4-1}
\usepackage{graphicx}
\usepackage{hyperref}
\usepackage{amsmath,amssymb}

\begin{document}	
\newcommand{\FIRSTAFF}{\affiliation{Department of Physics,
		University at Buffalo, SUNY
		Buffalo,
		NY 14260
		USA}}

\author{Wei-Chen Lin}
\email{weichenl@buffalo.edu} 
\author{Rance M. Solomon}
\email{rancesol@buffalo.edu}
\FIRSTAFF{}
\begin{abstract}
We study the general embedding of a $ P(X, \varphi) $ inflationary theory into a two-field theory with curved field space metric, which was proposed as a possible way to examine the relation between de Sitter Swampland conjecture and \textit{k}-inflation. We show that this embedding method fits into the special type of two-field model in which the heavy field can be integrated out at the full action level. However, this embedding is not exact due to the upper bound of the effective mass of the heavy field. We quantify the deviation between the speed of sound calculated via the $ P(X, \varphi) $ theory and the embedding two-field picture to next leading order terms. We especially focus on the first potential slow roll parameter defined in the two-field picture and obtain an upper bound on it.

\end{abstract}

\date{\today}

\title{Generalizing the Swampland: Embedding $P(X, \varphi)$ Inflationary Theories in a Curved Multi-field Space }

\maketitle
\section{Introduction}

In inflationary cosmology \cite{Starobinsky:1980te,Sato:1981ds,Sato:1980yn,Kazanas:1980tx,Guth:1980zm,Linde:1981mu,Albrecht:1982wi}, scalar fields have been extensively used for modeling the inflaton field causing an exponential expansion in the very early universe which sets up the initial condition of the Hot Big Bang. In the simplest scenario, a period of quasi-de Sitter phase is accomplished by a canonical scalar field slow-rolling down a plateau potential. The quantum fluctuations based on such a simple model are phenomenologically sufficient to match with the slightly tilted power spectrum and non-Gaussianity in the latest measurement of the anisotropies of the cosmic microwave background \cite{Akrami:2018odb, Akrami:2019izv}. Meanwhile, the supposed high energy scale during inflation gives rise to the question: what beyond standard model theory or quantum gravity theory contains the inflaton field? As the most prominent candidate of an ultraviolet (UV) complete theory of fundamental interactions, string theory is widely used to construct the inflaton field.

However, recently the existence of such a simple scalar field with a plateau potential from string theory construction has been put into question. Over the past several years, a series of so-called “swampland” conjectures have been proposed concerning the consistency of effective scalar field models with a UV completion in a theory of quantum gravity \cite{Vafa:2005ui}, including the de Sitter Swampland Conjecture and its modified versions \cite{Obied:2018sgi,Garg:2018reu,Ooguri:2018wrx}. The de Sitter Swampland conjecture states that for a scalar potential $V$ of any consistent theory
of quantum gravity, there is a lower bound on the logarithmic gradient of  $ V(\phi) $
\begin{equation}
\label{de_SitterSW}
M_P\frac{|V'(\phi)|}{V}\gtrsim c \sim\mathcal{O}(1).  
\end{equation}
While simple single-field inflation models are in strong tension with the swampland conjectures \cite{Agrawal:2018own,Kinney:2018nny}, more complicated models can in principle be consistent. In Ref. \cite{Achucarro:2018vey}, Ach\'ucarro and Palma  point out that in the multi-field models, the parameter characterizing the logarithmic gradient of the potential $ \epsilon_{V} $ can largely deviate from the 
first Hubble slow-roll parameter $ \epsilon $ through a large turning rate $ \Omega $ as
\begin{equation}
\label{epsilon_turning_mismatch*}
\epsilon_{V} \equiv \frac{M_P^2}{2} \frac{G^{IJ}\partial_{I}V\partial_{J}V}{V^2}   \simeq \epsilon\left(1+\frac{\Omega^2}{9H^2}\right).   
\end{equation}  
Therefore, the de Sitter Swampland conjecture $ \epsilon_{V} \gtrsim \mathcal{O}(1) $ can be satisfied without violating quasi-de Sitter condition $ \epsilon \ll 1 $. For instance, in hyperbolic inflation \cite{Brown:2017osf}, the inflaton can orbit a steep potential and never slow rolls.

Besides having more than one field, another way to go beyond simple inflationary models is by considering models based on the $ P(X, \varphi) $ theory \cite{ArmendarizPicon:1999rj, ArmendarizPicon:2000ah}, in which the Lagrangian is a general function of the field and its canonical kinetic term $ X=-(\partial \varphi)^2/2 $. For instance, the well-known Dirac-Born-Infeld (DBI) inflation \cite{Silverstein:2003hf, Alishahiha:2004eh} motivated by string theory construction belongs to this category. It is then a natural question to ask how to apply the Swampland conjectures on this type of model. Some possible methods are proposed aiming to generalize the conjectures themselves  
\cite{Seo:2018abc, Mizuno:2019pcm}. On the other hand, another rather conservative approach suggested in Ref. \cite{Dasgupta:2018rtp, Mizuno:2019pcm} and further studied in Ref. \cite{Solomon:2020viz} is to embed the $P(X, \varphi)$ theory into a two-field model\footnote{Other discussions related to the two-field models as (partial) UV completion can be found in Ref. \cite{Babichev:2017lrx, Babichev:2018twg, Mukohyama:2020lsu}.} by a general embedding scheme \cite{Elder:2014fea}
\begin{equation}
\label{embedding_k_inflation**}
S=\int d^{4}x\sqrt{-g}\left[\frac{M_P^2}{2}R+P(\varphi, \chi)-\frac{1}{2\Lambda^{6}}(\partial \chi)^2 +(X-\chi)P_{\chi}\right],
\end{equation}
where $\Lambda$ is a constant parameter with mass dimension.\footnote{Here, we adopt the notation used in Ref. \cite{Solomon:2020viz}.} When $\Lambda \rightarrow \infty$ the action reduces back to that of a $ P(X, \varphi) $ theory. In Ref. \cite{Solomon:2020viz}, Solomon and Trodden derive the first potential slow-roll parameter $\epsilon_V$ in the embedding two-field model as 
\begin{equation}
\label{Solomon_Trodden_result}
\epsilon_V \simeq  \epsilon \left[1  - \left(\frac{1-c_S^2}{1+c_S^2}\frac{\beta}{6}\right)+ \left(\frac{1-c_S^2}{1+c_S^2}\frac{\beta}{6}\right)^2\right]
\end{equation}
where $ \beta $ is a parameter defined in a $ P(X, \varphi) $ theory as
\begin{equation}
\label{beta*}
\beta \equiv \frac{\dot{\varphi}P_{\varphi X}}{HP_X}. 
\end{equation}
They conclude that to satisfy the de Sitter Swampland conjecture in this “UV extension” of the $ P(X, \varphi) $ theory, the condition is 
\begin{equation}
\label{Solomen_Trodden_condition}
\frac{1-c_S^2}{1+c_S^2} \beta > \mathcal{O}(\frac{1}{\sqrt{\epsilon}}) \gg 1.
\end{equation}

In this paper, we point out that the resulting two-field model of this general embedding method should be considered as a special type of two-field model with mass hierarchy, in which the heavy field can be integrated out at the full action level as was done in the \textit{gelaton} scenario \cite{Tolley:2009fg, Garcia-Saenz:2019njm}. As a result, the two-field extension has to satisfy the energy conditions where this embedding is valid. This requirement gives an upper bound of the parameter $ \Lambda $ in the action (\ref{embedding_k_inflation**}). Furthermore, we derive $ \epsilon_{V} $ in series expansion of $ \Lambda^6 $ and showing that the two $\beta$-dependent terms in Eq. (\ref{Solomon_Trodden_result}) belong to the next leading order.  Especially, we conclude that to have a small deviation between the speed of sound derived from the $P(X, \varphi)$ theory and the corresponding two-field model, those two terms must be small compared to a leading term proportional to $ \Lambda^6  $ in Eq. (\ref{final_epsilonV}).

The organization of this paper is as follows. In Sec. \ref{Secsub.P_theory} and  \ref{Secsub_multifield}, we briefly review the $ P(X, \varphi) $ theory and the multi-field inflationary models with curved field space. In Sec. \ref{SecSub_gelaton}, we review the gelaton scenario, a special type of multi-field model in which the heavy field can be integrated out at the full action level. In Sec. \ref{Sec_general_embed}, we show that the general embedding method (\ref{embedding_k_inflation**}) fits into the same category as the gelaton case. To the leading order, both the turning rate and effective mass of the entropic fluctuations are proportional to a single parameter $ \Lambda^6 $, which is the same parameter used in the action to turn off the kinetic term of the auxiliary field $\chi$. In Sec. \ref{Sec_de_Sitter}, we have our main results. We first show that to correctly relate $ \epsilon_{V} $ to the turning rate $ \Omega $, we need to include next to leading order terms in $\Lambda$. We  obtain $ \epsilon_{V} $ and the speed of sound $ c_S $ derived from the embedding two-field model in terms of quantities defined in $ P(X, \varphi) $ to the next leading order. From the condition that the two-field model and $ P(X, \varphi) $ should give similar speed of sound, we conclude that the $ \epsilon_{V} $ can only be dominated by the leading term proportional to $ \Lambda^6 $ in a large turning scenario. Lastly, an upper bound on $\epsilon_V$ of this type of model is given. The conclusion is in Sec. \ref{Sec_Conclusion}.

\section{$P(X, \varphi)$ theory and multi-field models with non-trivial field space metric }\label{Sec. background}

\subsection{$P(X, \varphi)$ theory}\label{Secsub.P_theory}

The action of a $ P(X, \varphi) $ theory is given by  
\begin{equation}
\label{k_inflation_action}
S=\int d^{4}x\sqrt{-g}\left[\frac{M_P^2}{2}R+P(\varphi, X)\right],
\end{equation}
where $ X=-\frac{1}{2}g^{\mu \nu}\partial_{\mu}\varphi\partial_{\nu}\varphi $ is the canonical kinetic term. We take a flat Friedmann-Robertson-Walker (FRW) metric,
\begin{equation}
\label{metric_-+++}
ds^2 =- dt^2 + a^2\left(t\right) d {\bf x}^2, 
\end{equation}
and consider only the homogeneous solutions of this scalar field. The energy density of the field is given by 
\begin{equation}
\label{energy density}
\rho(X,\varphi)= 2XP_X-P(X,\varphi),
\end{equation}
and the Friedmann and Raychaudhuri equations are
\begin{align}
\label{Friedmann}
H^2&=\frac{1}{3M_P^2}\rho=\frac{1}{3M_P^2}(2XP_X-P),
\\
\label{Raychaudhuri}
\frac{\ddot{a}}{a}&=-\frac{1}{6M_P^2}(\rho+3P)= -\frac{1}{3M_P^2}(XP_X +P),
\end{align}
where $ a(t) $ is the scale factor and $ H=\dot{a}/a $ is the Hubble parameter. Here, the subscript "X" indicates a derivative with respect to the canonical kinetic term, $ P_X \equiv \partial  P/\partial X $.  The resulting equation of motion (EoM) for the field $ \varphi $ is 
\begin{equation}
\label{EoM_non_canonical}
\ddot{\varphi}+3H\dot{\varphi}+\frac{\dot{\varphi}\dot{P_X}}{P_X}-\frac{P_{\varphi}}{P_{X}}=0.
\end{equation}
Besides the Hubble slow roll parameters quantifying the background evaluation 
\begin{equation}
\label{epsilon_eta}
\epsilon \equiv -\frac{\dot{H}}{H^2}=\frac{3}{2}(1+\frac{P}{\rho}), \qquad  \eta \equiv \frac{\dot{\epsilon}}{H\epsilon}, 
\end{equation}
it is useful to define a new parameter $ \tilde{\kappa} $ as
\begin{equation}
\label{kappa}
\tilde{\kappa} \equiv \frac{\dot{P_X}}{H P_X},   
\end{equation}
which quantifies the deviation from a canonical Lagrangian. To see this, we first notice that the first Hubble slow-roll parameter $ \epsilon $ in the single non-canonical field model can also be written as 
\begin{equation}
\label{epsilon_single_non_canon}
\epsilon=\frac{XP_X}{M_P^2H^2}.
\end{equation} 
Then by taking the time derivative of Eq.  (\ref{epsilon_single_non_canon}) and substituting Eqs. (\ref{epsilon_eta}, \ref{kappa}), we can derive the following relation
\begin{equation}
\label{EoM_non_can_in_parameters}
\ddot{\varphi}=\left(\frac{\eta}{2}-\epsilon-\frac{\tilde{\kappa}}{2}\right)H\dot{\varphi}, 
\end{equation}
which reduces back to the familiar relation of a canonical field in a FRW spacetime when $ \tilde{\kappa}=0 $.

In a $ P(X, \varphi) $ theory, it is well known that the perturbations travel with a speed of sound 
\begin{equation}
\label{SpeedofSound}
c_S^2=\frac{P_X}{P_X+2XP_{XX}}. 
\end{equation}
With Eq. (\ref{SpeedofSound}), we can further rewrite $ \tilde{\kappa} $ as
\begin{equation}
\label{new_slow_roll_para}
\tilde{\kappa} \equiv \frac{\dot{P_X}}{H P_X} =\left(\eta -2\epsilon \right)\left(\frac{1-c_S^2}{1+c_S^2}\right) +\frac{2\beta c_S^2}{1+c_S^2},  
\end{equation} 
where $ \beta $ is given by 
\begin{equation}
\label{beta}
\beta \equiv \frac{\dot{\varphi}P_{\varphi X}}{HP_X}, 
\end{equation}
is a useful parameter introduced in Ref. \cite{Solomon:2020viz}, and will be used in the later discussion. Notice that $ \beta $ is a parameter invariant from field redefinition in a $ P(X, \varphi) $ theory. Furthermore, $ P(X, \varphi) $ theories reduce to canonical limit only when both $ c_S \rightarrow 1 $ and $ \beta \rightarrow 0 $, which when substituted into Eq. (\ref{new_slow_roll_para}) give $ \tilde{\kappa} \rightarrow 0 $. Assuming $ \epsilon, \eta \ll 1$, we have
\begin{equation}
\label{kappa_slow_roll}
\tilde{\kappa} \simeq \frac{2\beta c_S^2}{1+c_S^2}. 
\end{equation}

\subsection{multi-field models with non-trivial field space metric}\label{Secsub_multifield}

Here we closely follow Refs. \cite{Achucarro:2018vey, Garcia-Saenz:2019njm} to review the subject of multi-field models with non-trivial field space metric.\footnote{Discussions related to integrating out the heavy degrees of freedom in multi-field models can be found in Refs. \cite{Chen:2009we, Achucarro:2010da, Achucarro:2010jv,Achucarro:2012yr,Cremonini:2010ua, Burgess:2012dz, Gwyn:2012mw, Gundhi:2018wyz, Gundhi:2020zvb}.} Consider a multi-field inflationary model described by the following non-linear sigma model
\begin{equation}
\label{non-linear-sigma model}
S=\int d^{4}x\sqrt{-g}\left[\frac{M_{P}^{2}}{2}R-\frac{1}{2}G_{IJ}(\phi) \partial^{\mu} \phi^{I} \partial_{\mu} \phi^{J} - V(\phi) \right], 
\end{equation}
where $ G_{IJ}(\phi) $ is the field space metric. Together with the FRW metric, the background dynamics is determined by the Friedmann equation
\begin{equation}
\label{MutliF_Friedmann}
H^2=\frac{1}{3M_P^2}\rho=\frac{1}{3M_P^2}\left(\frac{1}{2}\dot{\Phi}^{2}+V\right), 
\end{equation} 
where  $ \dot{\Phi}=\sqrt{G_{IJ}\dot{\phi}^{I}\dot{\phi}^{J}} $, and the EoM of the fields 
\begin{equation}
\label{MuiltF_EoM}
D_{t}\dot{\phi}^{I}+3H\dot{\phi}^{I}+G^{IJ}V_{J}=0,
\end{equation}
where $ V_{I} \equiv \partial V/ \partial \phi^I $ and $ D_{t} $ is the covariant time derivative with respect to the field metric, defined as $  D_{t}A^{I}= \dot{A}^{I}+\Gamma^{I}_{JK}\dot{\phi}^{J}A^{K}  $.   The first Hubble show-roll parameter $\epsilon $ (\ref{epsilon_eta}) is related to $ \dot{\Phi} $ as
\begin{equation}
\label{epsilon_two_field}
\epsilon=\frac{\dot{\Phi}^2}{2H^2M_P^2}.
\end{equation}
Taking the time derivative of Eq. (\ref{epsilon_two_field}) and using Eq. (\ref{epsilon_eta}), one can obtain a relation similar to that of a canonical single-field case
\begin{equation}
\label{Phi_DDot_multi}
\ddot{\Phi}=\left(\frac{\eta}{2}-\epsilon\right)H\dot{\Phi}.
\end{equation}
However, notice that $ \ddot{\Phi} $ includes the time derivative on the non-trivial field space metric since  $ \dot{\Phi}=\sqrt{G_{IJ}\dot{\phi}^{I}\dot{\phi}^{J}}$.

Next, in the two-field model, it is more useful to consider the field components tangent and normal to the trajectory in field space. For a given trajectory in the field space, the unit tangent vector is given by
\begin{equation}
\label{Fieldspace_tangentV}
T^{I}=\frac{\dot{\phi}^{I}}{\dot{\Phi}},
\end{equation}
and the unit normal vector (entropic direction) can be defined as
\begin{equation}
\label{Fieldspace_normalV}
N^{I}\equiv \frac{-D_{t}T^I}{|D_{t}T|} \equiv \frac{-D_{t}T^I}{\Omega},
\end{equation}
where $ \Omega \equiv |D_{t}T|  $ is the turning rate. By projecting onto the tangent and normal direction, the EoMs of the fields Eq. (\ref{MuiltF_EoM}) become
\begin{align}
\label{EOM_tangent}
&\ddot{\Phi}+3H\dot{\Phi}+V_{\Phi}=0,  \\
\label{EOM_normal}
&\Omega=\frac{V_{N}}{\dot{\Phi}},
\end{align}
where $ V_{\Phi} \equiv T^{I}V_{I}  $ and $ V_{N} \equiv N^{I}V_{I}$. Substituting Eq. (\ref{Phi_DDot_multi}) into Eq. (\ref{EOM_tangent}), we have 
\begin{equation}
\label{VPhi_in_slowPara}
\left(\frac{\eta}{2}-\epsilon+3\right)H\dot{\Phi}=-V_{\Phi}. 
\end{equation}
Due to the non-trivial field space metric $ G_{IJ}(\phi) $, the tangential direction differs from the direction defined by the gradient flow of the potential $ \partial_{I}V $, and as the result, slow roll limit can be achieved even with a steep potential. Quantitatively, this is given by the relation between the first Hubble slow-roll parameter $ \epsilon $ (\ref{epsilon_eta}) and the first potential Hubble parameter $ \epsilon_V $, 
\begin{equation}
\label{Multi_potential_epsilon}
\epsilon_{V} \equiv \frac{M_P^2}{2} \frac{V^{I}V_{I}}{V^2}=\frac{M_P^2}{2} \frac{V_{\Phi}^2+V_{N}^2}{V^2}. 
\end{equation}
Substituting Eqs. (\ref{EOM_normal}, \ref{VPhi_in_slowPara}) into Eq. (\ref{Multi_potential_epsilon}) and using Eqs. (\ref{MutliF_Friedmann}, \ref{epsilon_two_field}), we then have  
\begin{equation}
\label{epsilon_turning_mismatch}
\epsilon_{V} = \frac{\epsilon}{(3-\epsilon)^2}\left[\left(3-\epsilon+\frac{\eta}{2}\right)^2 +\frac{\Omega^2}{H^2}\right]   \simeq \epsilon\left(1+\frac{\Omega^2}{9H^2}\right).   
\end{equation}       
Notice that the first equality is an exact relation without using any approximation and in the second expression the general slow-roll condition $ \epsilon, \eta \ll 1 $ is assumed. So the de Sitter Swampland conjecture $ \epsilon_{V} \gtrsim \mathcal{O}(1) $ and the slow roll condition $ \epsilon \ll 1 $ can both be simultaneously satisfied if  $ \Omega \gg H $. Notice that although the action given by Eq. (\ref{non-linear-sigma model}) is non-canonical, the first potential Hubble parameter $ \epsilon_V $ (\ref{Multi_potential_epsilon}) is still well-defined, \textit{i.e.}, $ \epsilon_V $ is invariant under the local field redefinition. Therefore, the de Sitter Swampland conjecture $ \epsilon_{V} \sim \mathcal{O}(1) $ would not be changed by a simple field redefinition.   

Lastly, in two-field models, the effective mass of the isocurvature perturbations is given by
\begin{equation}
\label{EFF_ISO_Mass}
M_{eff}^2=V_{;NN}+\epsilon H^2 M_{P}^2 R_{fs}-\Omega^2,
\end{equation}
where $ V_{;NN}=N^{I}N^{J}V_{;IJ} $ is the projection of the covariant Hessian of the potential along the entropic direction, and $ R_{fs} $ is the Ricci scalar defined by the field space metric $ G_{IJ} $. If the effective mass of the isocurvature modes is heavy compared to the Hubble parameter, \textit{i.e.} $ M_{eff} \gg H $, then the heavy field (isocurvature modes) can be integrated out, resulting in an effective single field inflationary model with reduced speed of sound
\begin{equation}
\label{multiF_SoundSpeed}
c_S^{2}=\left(1+\frac{4\Omega^2}{M_{eff}^2}\right)^{-1}.
\end{equation} 
Notice that in general, this mapping is only valid at the perturbation level, since the heavy field is the isocurvature perturbations, not the field defined in the full action. However, for some special two-field models, the heavy field can be integrated out at the full action level \cite{Tolley:2009fg, Garcia-Saenz:2019njm}, in which assuming $ M_{eff} \rightarrow \infty $ is equivalent to turning off the dynamics of the auxiliary field. In the next subsection, we review the prototype of this type of model named the gelaton scenario.

\subsection{the gelaton scenario }\label{SecSub_gelaton}
Here, we review a special case of the gelaton scenario introduced by Tolley and Wyman \cite{Tolley:2009fg}, in which  by integrating out the heavy field, the effective low energy single field action is of the DBI form. Especially, we take the point of view from Ref. \cite{Garcia-Saenz:2019njm} that it can be considered as a special type of multi-field model with curved field space and mass hierarchy. Related discussion can be found in Refs. \cite{ Mizuno:2019pcm, Mukohyama:2020lsu}.  

Consider the following two-field action 
\begin{equation}
\label{gelaton_action}
S=\int d^{4}x\sqrt{-g}\left[-\frac{1}{2}(\partial \chi)^2 -\frac{e^{2g\chi/M_P}}{2}(\partial \varphi)^2-V(\chi, \varphi)\right],
\end{equation}  
with the potential
\begin{equation}
\label{DBI_EFT_potential}
V(\chi, \varphi)=T(\varphi)\left[\cosh(\frac{2g\chi}{M_{P}})-1\right]+U(\varphi), 
\end{equation}
where $ g $ is some dimensionless constant. The field $ \chi $ is called the gelaton field, which we will see later is the heavy field, and $ \varphi $ is the inflaton field.\footnote{Notice that the convention we use here is opposite to the convention in the original paper \cite{Tolley:2009fg}.} Comparing to Eq. (\ref{non-linear-sigma model}), one can immediately tell that Eq. (\ref{gelaton_action}) is a special case of a two-field non-linear sigma model with $ \phi^{I}=(\chi, \varphi) $ and field space metric given by 
\begin{equation}
\label{2D_hyperbolic_metric}
G_{IJ}(\phi)=diag(1, e^{2g\chi/M_P}). 
\end{equation}
From Eq. (\ref{gelaton_action}), the equation of motion of the field $ \chi $ is given by
\begin{equation}
\label{gelaton_DBI_EOM}
\Box\chi +\frac{g}{M_P}\left[e^{2g\chi/M_P}\dot{\varphi}^2-2T(\varphi)\sinh(\frac{2g\chi}{M_{P}})\right]=0.
\end{equation}
For large value of $ g $, we can drop the $ \Box\chi $ term and Eq. (\ref{gelaton_DBI_EOM}) reduces to a constraint on  $ \chi $ as 
\begin{equation}
\label{gelaton_constriant}
e^{2g\chi/M_P}\dot{\varphi}^2-2T(\varphi)\sinh(\frac{2g\chi}{M_{P}}) \simeq 0,  
\end{equation}
which can be further rewritten as 
\begin{equation}
\label{gelaton_constriant*}
e^{-4g\chi/M_P} \simeq \left(1-\frac{2X}{T(\varphi)}\right),  
\end{equation}
where $ X=-(\partial \varphi)^2/2 $.
We then can substitute Eq. (\ref{gelaton_constriant*}) into Eq. (\ref{gelaton_action}) and drop the kinetic term of the gelaton field to obtain the effective DBI action as
\begin{equation}
\label{DBI_action}
S_{eff}=\int d^{4}x\sqrt{-g} P(X,\varphi) =\int d^{4}x\sqrt{-g}\left[T(\varphi)\left(-\sqrt{1-\frac{2X}{T(\varphi)}}+1\right)-U(\varphi)\right].
\end{equation}
Notice that the speed of sound (\ref{SpeedofSound}) derived from the DBI action (\ref{DBI_action}) is inversely proportional to $  P_{X} $ as
\begin{equation}
\label{SpeedofSound_DBI}
c_S^2=P_{X}^{-2}=\left(1-\frac{2X}{T(\varphi)}\right).  
\end{equation}
Together with Eq. (\ref{gelaton_constriant*}), we can see that in the large $ g $ limit, the gelaton field is determined by the speed of sound or $  P_{X} $ as 
\begin{equation}
\label{gelaton_value}
\chi \simeq \frac{M_P}{2g}\ln \frac{1}{c_S}=\frac{M_P}{2g}\ln  P_{X} . 
\end{equation}

This gelaton scenario, in which the gelaton field can be integrated out at the full action level, can be considered as a special case of the more general multi-field models with heavy entropic fluctuations \cite{Garcia-Saenz:2019njm}. 
Therefore, Eq. (\ref{SpeedofSound_DBI}) should agree with Eq. (\ref{multiF_SoundSpeed}) under the same approximation, \textit{i.e.},
large $ g $ limit and dropping the kinetic term of $ \chi $ in the action (\ref{gelaton_action}).  
From the field space metric (\ref{2D_hyperbolic_metric}), we can derive the non-zero components of the Christoffel symbol and Ricci scalar as
\begin{equation}
\label{Christoffel_gelaton}
\Gamma^{\chi}_{\varphi\varphi}=\frac{g}{M_P}e^{-2g\chi/M_P}, \qquad  \Gamma^{\varphi}_{\chi\varphi}=\frac{g}{M_P},  \qquad R_{fs}=\frac{-2g^2}{M_P^2}.
\end{equation} 
In the large $ g $ limit, the unit tangent and normal vector Eqs. (\ref{Fieldspace_tangentV}, \ref{Fieldspace_normalV}) can be approximated by
\begin{equation}
\label{gelaton_T_N_vector_approx}
T^{I}=(0, e^{-g\chi/M_P} ),  \qquad  N^{I}=(-1, 0),  
\end{equation}  
and the resulting turning rate and the projection of the covariant Hessian of the potential along the entropic direction are given by 
\begin{equation}
\label{gelaton_turning}
\Omega^2=G_{IJ}D_{t}T^ID_{t}T^J \simeq\frac{g^2}{M_P^2}e^{2g\chi/M_P}\dot{\varphi}^2 
\end{equation}
and
\begin{equation}
\label{gelaton_Hessian}
V_{;NN}=N^{I}N^{J}V_{;IJ} \simeq \frac{4g^2}{M_P^2}T(\varphi)\cosh(\frac{2g\chi}{M_P} ),  
\end{equation} 
respectively. In the limit $ g \rightarrow \infty$, though the directions of the tangent and normal vectors are fixed in the $ (\chi, \varphi) $ basis, the turning rate of the trajectory $ \Omega \equiv |D_{t}T| $ is still nonzero due to the non-trivial field space metric (\ref{2D_hyperbolic_metric}). Next, by substituting the above results into Eq. (\ref{EFF_ISO_Mass}) with 
\begin{equation}
\label{H_dot}
-\dot{H}=\epsilon H^{2}=\frac{1}{2M_P^2}G_{IJ}\dot{\phi}^I\dot{\phi}^J \simeq \frac{1}{2M_P^2} e^{2g\chi/M_P}\dot{\varphi}^2,
\end{equation}
where the approximation is from the fact that the kinetic term of the gelaton field is sub-dominated, the effective mass of the isocurvature modes in the large $ g $ limit is given by
\begin{equation}
\label{gelaton_mass}
M_{eff}^2 \simeq \frac{2g^2}{M_P^2}\left[2T(\varphi)\cosh(\frac{2g\chi}{M_P})-e^{2g\chi/M_P}\dot{\varphi}^2\right]. 
\end{equation}
By substituting Eqs. (\ref{gelaton_turning}, \ref{gelaton_mass}) into Eq. (\ref{multiF_SoundSpeed}), we then have
\begin{equation}
\label{gelaton_speed_of_sound_1}
c_S^2 \simeq -1+\frac{4T(\varphi)}{2T(\varphi)+\left(1+\tanh(2g\chi/M_P)\right)\dot{\varphi}^2}  \simeq 1-\frac{2X}{T(\varphi)}, 
\end{equation}
where the constraint (\ref{gelaton_constriant*}) is used in the second approximation, and the result is just Eq. (\ref{SpeedofSound_DBI}). This shows that integrating out the gelaton field at the full action level is consistent with the picture of having heavy isocurvature modes. 
To gain a better understanding of the above quantities in the large $ g $ limit, we can use Eq. (\ref{gelaton_value}) to replace $ \chi $ in terms of $ c_S $ so that  

\begin{equation}
\label{gelaton_mass_&_turning}
M_{eff}^2 \simeq 4c_S\frac{g^2}{M_P^2}T(\varphi),  \qquad   \Omega^2 \simeq \frac{g^2}{M_P^2}\frac{1-c_S^2}{c_S}T(\varphi).
\end{equation}
We emphasize that both $ \Omega^2 $ and $ M_{eff}^2 $ are  proportional to $ g^{2} $ at the leading order, so in the large $ g $ limit, the speed of sound (\ref{multiF_SoundSpeed}) does not depend on $ g $. This limit is also a reasonable condition for dropping the kinetic term of $ \chi $ in the full action, and at the same time makes $ M_{eff} \gg H $, which is the condition for integrating out the heavy isocurvature modes. However, in the gelaton scenario, the effective mass is also bounded from above by the cutoff of the $ P(X, \varphi) $ theory in order to have a
weakly coupled perturbation theory, that is, the effective mass is bounded within the range 
\begin{equation}
\label{energy_condition}
H \ll M_{eff}<E.  
\end{equation}
With the resulting DBI model being an effective field theory of inflaton, which can be interpreted as the Goldstone boson corresponding to broken time translations \cite{Cheung:2007st}, the cutoff energy scale is given by 
\begin{equation}
\label{UV_cutoff}
E^4 \simeq 16\pi^2M_P^2 |\dot{H}|\frac{c_S^5}{1-c_S^2}.
\end{equation}
Since the effective mass of the gelaton (\ref{gelaton_mass_&_turning}) is bounded by the energy condition Eq. (\ref{energy_condition}), the value of $ g $ cannot be arbitrarily large. As a result, there is an $ \mathcal{O}(1/g^{2}) $ correction to the speed of sound given in Eq. (\ref{gelaton_speed_of_sound_1}). Notice that one can obtain the above results (\ref{gelaton_speed_of_sound_1}, \ref{gelaton_mass_&_turning}) without using approximation Eq. (\ref{gelaton_T_N_vector_approx}) by calculating exact results from Eqs. (\ref{Fieldspace_tangentV}, \ref{Fieldspace_normalV}) first, and then taking the leading order in  $ g $. 

In the next section we will see that the general embedding of a $ P(X,\phi) $ into a two-field non-linear sigma model also belongs to this category, \textit{i.e.}, making $ M_{eff} $ large and turning off the dynamics of the auxiliary field is controlled by a single parameter. Also we will use the method described above, since later in Sec. \ref{Sec_de_Sitter} we will see that the approximated tangent and normal vectors (\ref{gelaton_T_N_vector_approx}) are insufficient.

\section{the general embedding of an inflationary $ P(X, \varphi) $ theory}\label{Sec_general_embed}

In this section, we show that the general embedding of a $ P(X, \varphi) $ theory introduced in Ref. \cite{Elder:2014fea} as the generalization of the gelaton scenario fits into the same picture.
That is, the embedding can be considered as a special type of two-field model in which the heavy field can be integrate out at the full action level. To our knowledge, this is the first time in the literature that this is explicitly shown.

To begin with, one can rewrite the action of a $ P(X, \varphi) $ theory (\ref{k_inflation_action})  by introducing an auxiliary field $ \chi $ 
\begin{equation}
\label{lagrangian_multiplier}
S=\int d^{4}x\sqrt{-g}\left[\frac{M_P^2}{2}R+P(\varphi, X)+(X-\chi)P_{\chi}\right],
\end{equation}
in which the equation of motion of the auxiliary field is just a constraint equation, $ \chi=X $, providing $  P_{\chi\chi} \neq 0 $. Then by introducing a small kinetic term, the action becomes
\begin{equation}
\label{embedding_k_inflation}
S=\int d^{4}x\sqrt{-g}\left[\frac{M_P^2}{2}R+P(\varphi, \chi)-\frac{1}{2\Lambda^{6}}(\partial \chi)^2 +(X-\chi)P_{\chi}\right],
\end{equation}
which can be considered as a two-field non-linear sigma model (\ref{non-linear-sigma model}) with the field space metric
\begin{equation}
\label{embedding_metric}
G_{IJ}(\phi)=diag (\frac{1}{\Lambda^6}, P_{\chi}), 
\end{equation}
and a potential 
\begin{equation}
\label{embedding_potential}
V(\phi)=-P+\chi P_{\chi}. 
\end{equation}
Then we can follow the same logic used in the gelaton scenario. The equation of motion of $ \chi $ derived from the action (\ref{embedding_k_inflation}) is
\begin{equation}
\label{chi_EOM_general}
\frac{1}{\Lambda^6}\Box \chi + P_{\chi\chi}(X-\chi)=0,
\end{equation}
which effectively reduces to the constraint equation, $ \chi=X $ in the large $ \Lambda $ limit. And the action (\ref{embedding_k_inflation}) can be treated as a special type of two-field model with heavy isocurvature modes in which the heavy field can be integrated out at the full action level. In the following, we will show that the leading order term of both $M_{eff}^2$ and $\Omega^2$ are proportional to $\Lambda^6$, so in the large $ \Lambda $ limit, the speed of sound given by Eq. (\ref{multiF_SoundSpeed}) matches the result given by the usual single field formula
\begin{equation}
\label{SpeedofSound*}
c_S^2=\frac{P_X}{P_X+2XP_{XX}}. 
\end{equation}

Firstly, the non-zero components of the Christoffel symbol derived from the field space metric (\ref{embedding_metric}) are: 
\begin{equation}
\label{Christoffel_embedding}
\Gamma^{\chi}_{\varphi\varphi}=\frac{-1}{2}\Lambda^6 P_{\chi\chi}, \qquad  \Gamma^{\varphi}_{\chi\varphi}=\frac{1}{2} \frac{P_{\chi\chi}}{P_{\chi}}, \qquad
\Gamma^{\varphi}_{\varphi\varphi}=\frac{1}{2} \frac{P_{\chi\varphi}}{P_{\chi}}.
\end{equation} 
And the Ricci Scalar of the field space is given by
\begin{equation}
\label{R_fs_embedding}
R_{fs}(\chi, \varphi)=\Lambda^6 \frac{\left[P_{\chi\chi}^2-2P_{\chi}P_{\chi\chi\chi}\right]}{2P_{\chi}^2},  
\end{equation} 
which is in general a function of the fields. Substituting the DBI Lagrangian (\ref{DBI_action}) into Eq. (\ref{R_fs_embedding}) with $ \chi \simeq X $, the Ricci scalar of the field space is 
\begin{equation}
\label{Ricci_embedding_DBI}
R_{fs}(\chi \simeq X, \varphi) \simeq \frac{-5\Lambda^6}{2(T(\varphi)-2X)^2}=\frac{-5 T(\varphi)^2}{2c_S^4}\Lambda^6,
\end{equation}
which is not a constant. Since the Ricci scalar is invariant from field re-definition, this general embedding method gives a different two-field picture compared to the previous gelaton scenario. Next, the unit tangent vector (\ref{Fieldspace_tangentV}) is given by 
\begin{equation}
\label{tangent_vector_embedding}
(T^{\chi}, T^{\varphi})=\frac{1}{\sqrt{\dot{\chi}^2/\Lambda^6+\dot{\varphi}^2P_{\chi}}}(\dot{\chi},\dot{\varphi}), 
\end{equation}
and for the two-field models, the corresponding unit normal vector (\ref{Fieldspace_normalV}) 
\begin{equation}
\label{normal_vector_embedding}
(N^{\chi}, N^{\varphi})=\frac{\sqrt{P_{\chi}/\Lambda^6}}{\sqrt{\dot{\chi}^2/\Lambda^6+\dot{\varphi}^2P_{\chi}}}\left(-\Lambda^6\dot{\varphi},\frac{\dot{\chi}}{P_{\chi}}\right),
\end{equation}
can be derived instead from the formula
\begin{equation}
\label{normal_vector_derivation}
N^{I}=-G^{IJ}\sqrt{\det G}\epsilon_{JK}T^{K},
\end{equation}
where $ \epsilon_{JK} $ is the Levi-Civita symbol. In the large $ \Lambda $ limit, taking only the leading order term in $ \mathcal{O}(\Lambda^6) $, the turning rate is given by
\begin{equation}
\label{embedding_turning}
\Omega^2=G_{IJ}D_{t}T^ID_{t}T^J \simeq \Lambda^6 \frac{\dot{\varphi}^2P_{\chi\chi}^2}{4P_{\chi}}, 
\end{equation}
and the projection of the covariant Hessian of the potential along the entropic direction is
\begin{equation}
\label{embedding_Hessian}
V_{;NN}=N^{I}N^{J}V_{;IJ} \simeq  \Lambda^6 \left(P_{\chi\chi}+\chi P_{\chi\chi\chi}\right). 
\end{equation}
Similar to the gelaton case, the above leading order results of $ \Omega^2 $ and $V_{;NN}$ can be derived by using the approximated unit tangent and normal vectors 
\begin{equation}
\label{tangent/normal_limit}
(T^{\chi}, T^{\varphi})\simeq\frac{1}{\sqrt{P_{\chi}}}\left(\frac{\dot{\chi}}{\dot{\varphi}},0\right), \qquad (N^{\chi}, N^{\varphi}) \simeq \Lambda^3\left(-1, \frac{1}{\Lambda^6 P_{\chi}}\frac{\dot{\chi}}{\dot{\varphi}}\right)\simeq \left(-\Lambda^3, 0\right), 
\end{equation}
and dropping the kinetic term of auxiliary field\footnote{This is not always true since there is an upper bound for $ \Lambda^6 $, Eq. (\ref{bounds_on_Lambda}). We will discuss the condition of this approximation in the end of this section. } 
\begin{equation}
\label{large_Lambda_approx_vectors}
\dot{\Phi}^2=\dot{\chi}^2/\Lambda^6 + \dot{\varphi}^2P_{\chi} \simeq \dot{\varphi}^2P_{\chi}.    
\end{equation}
Substituting Eqs. (\ref{R_fs_embedding}, \ref{embedding_turning}, \ref{embedding_Hessian}) into Eq. (\ref{EFF_ISO_Mass}) with 
\begin{equation}
\label{H_dot_embedding}
-\dot{H}=\epsilon H^{2}=\frac{1}{2M_P^2}G_{IJ}\dot{\phi}^I\dot{\phi}^J \simeq \frac{1}{2M_P^2} P_{\chi}\dot{\varphi}^2,
\end{equation}
we have the effective mass of the isocurvature modes as
\begin{equation}
\label{embedding_ISO_mass}
M_{eff}^2 \simeq  \Lambda^6\frac{P_{\chi\chi}}{4}\left(4+\frac{(2\chi-\dot{\varphi}^2)P_{\chi\chi}}{P_{\chi}}\right).
\end{equation}
Substituting Eq. (\ref{embedding_turning}, \ref{embedding_ISO_mass}) into Eq. (\ref{multiF_SoundSpeed}), we then have
\begin{equation}
\label{SoundSpeed_embedding}
c_S^2 \simeq \frac{4P_{\chi}+(2\chi-\dot{\varphi}^2)P_{\chi\chi}}{4P_{\chi}+(2\chi-\dot{\varphi}^2)P_{\chi\chi}+4\dot{\varphi}^2P_{\chi\chi}}, 
\end{equation}
which reduces back to the usual form of speed of sound of a $ P(X, \varphi) $ theory (\ref{SpeedofSound*}) when $ \chi=\dot{\varphi}^2/2  $. Therefore, we conclude that this general embedding scheme is similar to the gelaton case in the sense that in the large $ \Lambda $ limit, the leading order term in $ \Omega^2 $ and $ M_{eff}^2 $ are proportional to $ \Lambda^6 $, and as a result the speed of sound (\ref{multiF_SoundSpeed}) is independent of $ \Lambda $. It is also in this same limit which we can drop the kinetic term of the auxiliary field $ \chi $, such that the EoM of it reduces to a constraint equation.  Substituting the constraint $ \chi=\dot{\varphi}^2/2 \equiv X $ into the above results, we have 
\begin{equation}
\label{embedding_results}
\Omega^2 \simeq \Lambda^6 \frac{XP_{XX}^2}{2P_{X}}, \qquad  M_{eff}^2 \simeq  \Lambda^6 P_{XX},  \qquad  H^2 \simeq \frac{XP_X}{M_P^2 \epsilon},  \qquad c_S^2 \simeq \frac{P_{X}}{P_{X}+2XP_{XX}},
\end{equation}
and notice that the quantities are related by 
\begin{equation}
\label{the_relation}
\Omega^2 \simeq  M_{eff}^2\frac{1-c_S^2}{4c_S^2}, 
\end{equation}
which is the consistency condition that Eq. (\ref{multiF_SoundSpeed}) and Eq. (\ref{SpeedofSound*}) should give the same result at the leading order approximation. However, it also shows that when $ c_S \approx 1 $, the mapping between the $ P(X, \varphi) $ theory and the corresponding two-field model is ill-defined. It is easy to understand from $  M_{eff}^2  $ and $ c_S^2 $ in Eq. (\ref{embedding_results}). When $ c_S \rightarrow 1 $, we have $  M_{eff}^2 \rightarrow 0  $, which indicates the break down of the mass hierarchy, $ H \ll M_{eff} $. Therefore, both of the degrees of freedom in the field space are relevant, and cannot be described by an effective $ P(X, \varphi) $ theory. In the rest of the paper, we do not consider $ c_S \approx 1 $. Lastly, using  Eqs. (\ref{embedding_potential}, \ref{large_Lambda_approx_vectors}) with $ \chi \simeq X $, one can show that the EoM of $ \Phi $ from the two-field picture Eq. (\ref{EOM_tangent}) reduces to the EoM of the $ P(X, \varphi) $ theory Eq. (\ref{EoM_non_canonical}). Meanwhile, under the same approximation the relation of the slow-roll parameters in the two-field models, Eq. (\ref{Phi_DDot_multi}), 
also reduces to the corresponding relation in a $ P(X, \varphi) $ theory, Eq. (\ref{EoM_non_can_in_parameters}), 
so the classical dynamics of the field $ \Phi $ and the background evolution are well described by the effective $ P(X, \varphi) $ theory. 
 
Since we've confirmed that for a $ P(X, \varphi) $ theory it can be viewed as the low energy EFT of a special type of two-field model (\ref{embedding_k_inflation}) with large $ \Lambda $, then, to be an EFT of inflation, the energy bound on the range of validity, Eqs. (\ref{energy_condition}, \ref{UV_cutoff}), should apply too. In the remainder of the article when referring to a $P(X,\varphi)$ theory, we implicitly consider only those with $\varphi$ being the inflationary field. Substituting the above result into  Eqs. (\ref{energy_condition}, \ref{UV_cutoff}), the bound on $ M_{eff}^2 $ for this general embedding scheme to be valid is given by 
\begin{equation}
\label{bounds_on_M_eff}
\frac{XP_X}{M_P^2 \epsilon} \ll  M_{eff}^2   <  4\pi \sqrt{XP_X\frac{c_S^5}{1-c_S^2}}, 
\end{equation}
which can be rewritten as a bound on $ \Lambda^6 $ as
\begin{equation}
\label{bounds_on_Lambda}
\frac{2X^2}{M_P^2 \epsilon}\frac{c_S^2}{1-c_S^2} \ll  \Lambda^6   < \frac{4\sqrt{2}\pi X}{\sqrt{P_{XX}}}\frac{c_S^{7/2}}{1-c_S^2}. 
\end{equation}
The first inequality indicates the condition that a model has to satisfy in order to integrate out the auxiliary field, \textit{i.e.}, large $ \Lambda $ limit, and the second inequality is the upper bound to avoid strong coupling issues of the $ P(X, \varphi) $ theory itself. That is, for consistency, we cannot embed a $ P(X, \varphi) $ theory in a two-field UV extension in which the effective mass of isocurvature modes is heavier than the cutoff energy of the  $ P(X, \varphi) $ theory. From the condition that the upper bound in Eq. (\ref{bounds_on_Lambda}) should be much greater than the lower bound, we immediately have a consistency relation 
\begin{equation}
\label{first_consistency}
H \ll \frac{4\pi c_S^{5/2}}{\sqrt{1-c_S^2}}\sqrt{\epsilon}M_P.
\end{equation}
Next, by using Eqs. (\ref{EoM_non_can_in_parameters}, \ref{embedding_results}), the large $ \Lambda $ approximation (\ref{large_Lambda_approx_vectors}) is valid when
\begin{equation}
\label{condtion of approximation}
\Lambda^{6} \gg \frac{2X^2}{M_P^2\epsilon}\left(\frac{\eta}{2}-\epsilon-\frac{\tilde{\kappa}}{2}\right)^2. 
\end{equation}
Compared to the lower bound in Eq. (\ref{bounds_on_Lambda}), the condition (\ref{large_Lambda_approx_vectors}) is satisfied  by any value of $ \Lambda^6 $ inside the range (\ref{bounds_on_Lambda}) if
\begin{equation}
\label{Compare_lowerB}
1 > \frac{(1-c_S^{2})}{c_S^2}\left(\frac{\eta}{2}-\epsilon-\frac{\tilde{\kappa}}{2}\right)^2.
\end{equation}
Notice that violation of this condition just means that the lower bound in Eq. (\ref{bounds_on_Lambda}) should be replaced by Eq. (\ref{condtion of approximation}), \textit{i.e.} one should use the stronger lower bound between Eqs. (\ref{bounds_on_Lambda}) and (\ref{condtion of approximation}).

\section{Next to the leading order calculation and $ \epsilon_{V} $ }\label{Sec_de_Sitter}

Due to the upper bound of $\Lambda$, the embedding picture is not exact. In this section we quantify the effect of a finite $\Lambda$ on the embedding picture.

First, in the two-field picture, given a field space metric and potential we are able to calculate the first potential slow-roll parameter from Eq. (\ref{Multi_potential_epsilon}) as
\begin{equation} \label{epsilon_V_embedding}
\epsilon_V = \frac{M_P^2}{2}\frac{\Lambda^6 \chi^2P_{\chi \chi}^2+\left(P_{\varphi}-\chi P_{\varphi \chi}\right)^2/P_{\chi}}{\left(\chi P_{\chi}-P\right)^2}. 
\end{equation}
Since $\Lambda$ is bounded from above in Eq. (\ref{bounds_on_Lambda}), $\epsilon_V$ cannot be made arbitrarily large by a choice of $\Lambda$. Moreover, we will find that the leading order terms in $V_N^2$ is  $\mathcal{O}(\Lambda^6)$ while $V_\Phi^2$ is only $\mathcal{O}(\Lambda^0)$. For this reason we will need to expand $V_N^2$ out to the next to leading order term in $\Lambda$ to see the full effect, which means the approximation of the unit tangent and normal vectors (\ref{tangent/normal_limit}) are insufficient. In the following, the strategy is to expand quantities in the two-field picture out to $ \mathcal{O}(\Lambda^0) $. Then, we express the relevant parameters in the two-field picture with those in the single-field $P(X,\varphi)$ theory by using the constraint with $ \mathcal{O}(1/\Lambda^6) $ correction derived from the EoM of the auxiliary field  (\ref{chi_EOM_general}) as\footnote{Notice that we assume that the quantum correction to the field  $ \chi $ is small compared to the classical $ \mathcal{O}(1/\Lambda^6) $ correction. }
\begin{equation}
\label{chi=X_approx}
\chi=X-\frac{1}{\Lambda^6}\frac{\Box\chi}{P_{\chi\chi}} \simeq X-\frac{1}{\Lambda^6}\frac{\Box X}{P_{XX}}. 
\end{equation} 
For instance, substituting this modified constraint into $ P_{\chi \chi} $ gives the following $ \mathcal{O}(1/\Lambda^6) $ correction   
\begin{equation}
\label{example_NL}
P_{\chi\chi}(\chi \rightarrow X-\frac{1}{\Lambda^6}\frac{\Box X}{P_{XX}}, \varphi ) \simeq P_{XX} +\left(-\frac{1}{\Lambda^6}\frac{\Box X}{P_{XX}}\right)P_{XXX}.
\end{equation}

To $\mathcal{O}(\Lambda^0)$, we can have 

\begin{equation} \label{VPhi^2}
V^2_{\Phi} \equiv (T^IV_I)^2 \simeq \frac{1}{\dot{\varphi}^2 P_{\chi}}\left[\dot{\varphi}\left(-P_{\varphi}+\chi P_{\varphi \chi}\right)+\chi \dot{\chi}P_{\chi \chi}\right]^2
\end{equation}
and 
\begin{equation} \label{VN^2}
V^2_{N}  \equiv (N^IV_I)^2 \simeq \Lambda^6 \chi^2 P_{\chi \chi}^2 - \frac{1}{\dot{\varphi}^2 P_{\chi}}\left[\chi \dot{\chi}P_{\chi \chi}\left(2\dot{\varphi}\left(-P_{\varphi}+\chi P_{\varphi \chi}\right)+\chi \dot{\chi}P_{\chi \chi}\right)\right],
\end{equation}
which when combined give
\begin{equation} \label{VPhi^2+VN^2}
V^2_{\Phi}+V^2_{N}=\Lambda^6 \chi^2P_{\chi \chi}^2+\left(P_{\varphi}-\chi P_{\varphi \chi}\right)^2/P_{\chi},
\end{equation}
as expected from Eq. (\ref{epsilon_V_embedding}).
Now, to evaluate $ V_{\Phi}^2 $, $ V_N^2 $ and $\epsilon_V$ in terms of quantities defined in a $ P(X, \varphi) $ theory, we first substitute Eq. (\ref{chi=X_approx}) into Eqs. (\ref{VPhi^2}, \ref{VN^2}, \ref{epsilon_V_embedding}) then keep terms to $\mathcal{O}(\Lambda^0) $ and rewrite them by using the $ P(X, \varphi) $ relations in Sec. (\ref{Secsub.P_theory}). 
Firstly, the two-field potential derivatives can be expressed in terms of the single-field $P(X,\varphi)$ parameters,
\begin{equation} \label{VPhi^2_in_parameters}
V^2_{\Phi} \simeq 2\epsilon M_P^2 H^4\left(3-\epsilon+\frac{\eta}{2}\right)^2
\end{equation}
and
\begin{equation} \label{VN_in_parameters}
V_{N}^2 \simeq \Lambda^6  P_{XX} \frac{1-c_S^2}{2c_S^2}\epsilon H^2 M_P^2 + 2\epsilon M_P^2 H^4 \left(\frac{\tilde{\kappa}}{2}-\frac{\beta}{2}\right) \left(6-2\epsilon +\eta +\frac{\tilde{\kappa}}{2}-\frac{\beta}{2}\right) + 4H^2 \Delta \left(X^2P_{XX}+X^3P_{XXX}\right),
\end{equation}
where $ \Delta $ is given by\footnote{The derivation is given in Eq. (\ref{Box_X}).} 
\begin{equation}
\label{Delta}
\Delta \equiv (\frac{\kappa \eta}{2}-\epsilon \eta -\frac{\tilde{\kappa} \tilde{\lambda}}{2})+(\frac{ \eta}{2}-\epsilon -\frac{\tilde{\kappa}}{2})(3+ \eta-3\epsilon -\tilde{\kappa}), 
\end{equation}
with higher order slow-roll parameters $ \kappa \equiv \dot{\eta}/(H\eta) $ and $ \tilde{\lambda} \equiv  \dot{\tilde{\kappa}}/(H\tilde{\kappa}) $. 
Continuing with $\mathcal{O}(\Lambda^0)$, $\epsilon_V$ is given by
\begin{equation} \label{epsilon_V_embedding_chi=X}
\epsilon_V  \simeq \frac{M_P^2}{2}\frac{\Lambda^6 X^2P_{XX}^2+\left(P_{\varphi}-XP_{\varphi X}\right)^2/P_{X}-2\Box X\left(XP_{XX}+X^2P_{XXX}+X^{3}P_{XX}^2/(XP_{X}-P)\right)}{\left(XP_{X}-P\right)^2}. 
\end{equation}
Notice that this result is different from the result in Ref. \cite{Solomon:2020viz} since the authors use only the leading order result of the constraint equation $ \chi \simeq X $. The additional terms $ -2\Box X\left(XP_{XX}+X^2P_{XXX}+X^{3}P_{XX}^2/(XP_{X}-P)\right) $ are from the $ \mathcal{O}(1/\Lambda^6) $ correction to the constraint equation (\ref{chi=X_approx}).\footnote{ $ -2\Box X\left(XP_{XX}+X^2P_{XXX}\right) $ is the same factor in $ V_N^2 $ Eq. (\ref{VN_in_parameters}), but $ -2\Box X\left(X^{3}P_{XX}^2/(XP_{X}-P)\right) $ is the factor from the combination of $\mathcal{O}(\Lambda^6)$ in $ V_N^2 $ and  $\mathcal{O}(1/\Lambda^6)$ correction to $ V(\chi, \varphi)^{-2} $ by substituting Eq. (\ref{chi=X_approx}). } Using the relations of the $ P(X, \varphi) $ theory to rewrite the terms in Eq. (\ref{epsilon_V_embedding_chi=X}), $\epsilon_V$ takes the form
\begin{equation} \label{epsilonV_result}
\epsilon_{V}  \simeq \frac{\epsilon}{(3-\epsilon)^2}\left[\frac{\Lambda^6  P_{XX}}{4 H^2} \frac{1-c_S^2}{c_S^2} + \left(3-\epsilon +\frac{\eta}{2} +\frac{\tilde{\kappa}}{2}-\frac{\beta}{2} \right)^2 + \Delta\left(\frac{3\lambda}{c_S^2\Sigma}-\frac{1-c_S^2}{2c_S^2}+ \frac{\epsilon}{3-\epsilon}\left(\frac{1-c_S^2}{2c_S^2}\right)^2\right) \right], 
\end{equation}
where $ \lambda, \Sigma $ are two parameters commonly used in the calculation of non-Gaussianity   
\begin{equation}
\label{lambda_non_Gaussianity}
\lambda \equiv X^2P_{XX}+\frac{2}{3}X^3P_{XXX},  \qquad \Sigma \equiv   XP_{X}+2X^2P_{XX},  
\end{equation}
and $ \Lambda^6 $ satisfying the bound (\ref{bounds_on_Lambda})
\begin{equation} \label{bounds_on_Lambda*}
\frac{2X^2}{M_P^2 \epsilon}\frac{c_S^2}{1-c_S^2} \ll  \Lambda^6   < \frac{4\sqrt{2}\pi X}{\sqrt{P_{XX}}}\frac{c_S^{7/2}}{1-c_S^2}. 
\end{equation}
Notice that there is no simple way to rewrite $ \lambda /\Sigma $ into combination of flow parameters and it must be treated case by case \cite{Chen:2006nt}. Also notice that the turning rate is related to $ V_N $ through the EoM along the normal direction Eq. (\ref{EOM_normal}), so the following relation is exact to any order,   
 \begin{equation}
 \label{VN_turning}
 V_{N}^2=\Omega^2 \dot{\Phi}^2 = 2\Omega^2 \epsilon H^2 M_P^2. 
 \end{equation}
 Substituting Eqs. (\ref{VPhi^2_in_parameters}, \ref{VN_turning}) into the definition of $ \epsilon_V $, (\ref{Multi_potential_epsilon}),  we then have 
 \begin{equation}
 \label{show_relation_approx}
 \epsilon_{V} =\frac{M_P^2}{2} \frac{V_{\Phi}^2+V_{N}^2}{V^2} \simeq \frac{\epsilon}{(3-\epsilon)^2}\left[\left(3-\epsilon+\frac{\eta}{2}\right)^2 +\frac{\Omega^2}{H^2}\right], 
 \end{equation}
 which is $ \epsilon_V $ in terms of the slow-roll parameters and turning rate. We emphasize that, although $\epsilon_V$ is exactly given by Eq. (\ref{epsilon_V_embedding}), to relate the deviation between $\epsilon$ and $\epsilon_V$ to the turning rate, we need to rewrite it into a form that is a combination of $V_\Phi^2$ and $V_N^2$. However, $V_N^2$ is a quantity with leading order in $\mathcal{O}(\Lambda^6)$ while $V_\Phi^2$ is only in  $\mathcal{O}(\Lambda^0)$. To have the correct approximation result, Eq. (\ref{show_relation_approx}), we need to include $\mathcal{O}(\Lambda^0)$. This would conclude the argument why the leading order approximation used in Sec. \ref{Sec_general_embed} is not enough.

Under the general slow-roll approximation $ \epsilon, \eta \ll 1$ and all higher order flow parameters assumed small except $ \beta $, $ \epsilon_{V} $ (\ref{epsilonV_result}) reduces to  
\begin{align}
\label{final_epsilonV}
\epsilon_{V}  
& \simeq  \epsilon \left[1 +\frac{\Lambda^6  P_{XX}}{36 H^2} \frac{1-c_S^2}{c_S^2} -  \left(\frac{1-c_S^2}{1+c_S^2}\frac{\beta}{3}\right)+ \left(\frac{1-c_S^2}{1+c_S^2}\frac{\beta}{6}\right)^2 + \frac{1}{9}\left(\frac{2 \beta^2 c_S^2}{(1+c_S^2)^2}-\frac{3 \beta}{(1+c_S^2)}\right)\left(\frac{3\lambda}{\Sigma}-\frac{1-c_S^2}{2}+ \frac{\epsilon(1-c_S^2)^2}{12c_S^2}\right) \right],   
\end{align}
where Eq. (\ref{kappa_slow_roll}) is used.
Further assuming $ c_S^2 \ll 1 $, we have
\begin{equation} 
\label{final_final_epsilonV}
\epsilon_V \simeq \epsilon \left[1 +\frac{\Lambda^6  P_{XX}}{36 H^2} \frac{1}{c_S^2} -  \frac{\beta}{3}+ \left(\frac{\beta}{6}\right)^2+ \frac{1}{9}(2\beta^2 c_S^2 -3 \beta)(\frac{3\lambda}{\Sigma}-\frac{1}{2}+\frac{\epsilon}{12c_S^2})\right].
\end{equation}
In Eq. (\ref{final_epsilonV}), the second term is from the leading order term in $ V_N^2 $ and the third to fifth terms include the next to the leading order contribution. Notice that the third and fourth terms are the two $ \beta $-dependent terms in Eq. (\ref{Solomon_Trodden_result}) found in Ref. \cite{Solomon:2020viz}. Since it is an order expansion, we argue that the second term must dominate in a large turning scenario. Especially, if the next to the leading order contribution is comparable to the size of second term, one should expect a significant modification to the speed of sound given by Eq. (\ref{SoundSpeed_embedding}), which is the result from the leading order approximation. If the two-field picture and the $ P(X, \varphi) $ theory give significantly different speeds of sound, then the whole embedding picture is no longer valid. A detailed calculation of the speed of sound (\ref{multiF_SoundSpeed}) to the next leading order is given in the Appendix \ref{Appendix}. When the speed of sound is small\footnote{To have the simple form in Eq. (\ref{SpeedofSound_NL_Approx}), we need to assume some other flow parameters are also small, which are discussed in the Appendix \ref{Appendix}. }, we can relate the two-field speed of sound $c_{s(2-field)}$ calculated from Eq. (\ref{multiF_SoundSpeed}) and the $P(X,\varphi)$ speed of sound $c_s$ calculated from Eq. (\ref{SpeedofSound}) by the following relation
\begin{equation} \label{SpeedofSound_NL_Approx}
c_{s(2-field)}^{-2} \simeq c_S^{-2}-\frac{6XH^2}{\Lambda^6 P_X}\left(2\beta +  \beta^2 c_S^2 \right).
\end{equation}
The requirement that the difference between $ c_{s(2-field)} $ and $ c_{s} $ is small leads to the following condition 
\begin{equation}
\label{SpeedSounds_small_difference}
\Lambda^6 \gg \frac{6XH^2}{P_X} |2\beta c_S^2 + \beta^2 c_S^4 |. 
\end{equation}
Substituting Eq. (\ref{SpeedSounds_small_difference}) into the $ \mathcal{O}(\Lambda^6) $ term of $ \epsilon_{V} $ (\ref{final_final_epsilonV}) we have 
\begin{equation}
\label{leading_term_size_in_beta}
\frac{\Lambda^6  P_{XX}}{36 H^2} \frac{1}{c_S^2} \gg \frac{1}{12}\left|\frac{2\beta}{c_S^2} + \beta^2 \right|.
\end{equation}
For a DBI model, 
\begin{equation}
\label{DBI_lambda/Sigma}
\frac{\lambda}{\Sigma} =\frac{X}{T(\varphi)-2X}=\frac{1-c_S^2}{2c_S^2},
\end{equation}
and Eq. (\ref{final_final_epsilonV}) becomes 
\begin{equation}
\label{DBI_epsilonV}
\epsilon_V \simeq \epsilon \left[1 +\frac{\Lambda^6  P_{XX}}{36 H^2} \frac{1}{c_S^2} -  \frac{\beta}{3}+ \left(\frac{\beta}{6}\right)^2+ \left(\frac{\beta^2}{3}-\frac{\beta}{3c_S^2}\right) \right],
\end{equation}
in which the only possible dominating $ \mathcal{O}(\Lambda^0) $ terms are terms proportional to $ \beta^2 $ or $ \beta/c_S^2 $. However, from Eq. (\ref{leading_term_size_in_beta}) we see that both terms must be small compared to the leading order term to satisfy the condition that the deviation of the speed of sound given by two-field model and $ P(X, \varphi) $ theory is small.  Though the term $ \lambda/\Sigma $ has different form for different $ P(X, \varphi) $ theories, to be consistent, the leading order term should dominate the sub-leading terms if the $ P(X, \varphi) $ theory is an effective field theory of the two-field action. That is, the deviation between $ \epsilon_{V} $ and $ \epsilon $ in this type of model should be dominated by the $ \mathcal{O}(\Lambda^6) $ term as
\begin{equation}
\label{DBI_epsilonV_LeadingOnly}
\epsilon_V \simeq \epsilon \frac{\Lambda^6  P_{XX}}{36 H^2} \frac{1}{c_S^2}.
\end{equation}

Though above analytic calculation is performed under the condition $ c_S \ll 1  $, one can use Eq. (\ref{SpeedofSound_NL_in_A_2}), which is the form valid for $ 0< c_S <1 $, to obtain the same conclusion that to have small next order correction to the speed of sound, the $ \mathcal{O}(\Lambda^6) $ term in $ \epsilon_{V} $ should be the dominating term in a large turning scenario.  The upper bound of $ \Lambda^6 $ is of particular interest since it gives the maximum deviation between $ \epsilon_{V} $ and $ \epsilon $. By using the upper bound in Eq. (\ref{bounds_on_Lambda}) and take only the leading contribution in Eq. (\ref{final_epsilonV}), we obtain the upper bound of $ \epsilon_{V} $ as
\begin{equation} \label{Bound_on_epsilon_V}
\epsilon_{V} < \frac{\pi}{9}\sqrt{(1-c_S^2)c_S}\epsilon^{3/2}\frac{M_P}{H}, 
\end{equation}
where $ \epsilon, \eta \ll 1$ is assumed and notice that $ H $ has to satisfy Eq. (\ref{first_consistency}). Then, for a given background evolution, we can determine how large $ \epsilon_{V} $ can be in this special UV extension. Next, for a $ P(X, \varphi) $ inflationary theory to satisfy the de Sitter Swampland conjecture in this UV extension, using the simplified version $ \epsilon_V >1 $, we then have an upper bound on the Hubble parameter as 
\begin{equation}
\label{Bound_on_H_UV}
H < \frac{\pi}{9}\sqrt{(1-c_S^2)c_S}\epsilon^{3/2}M_P.
\end{equation} 
Notice that a choice of $  P(X,\varphi) $ theory must satisfy Eq. (\ref{first_consistency}) in order for this embedding method to work. The model must also satisfy Eq. (\ref{Bound_on_H_UV}) for the de Sitter conjecture to be met in this embedding.

Interestingly, another recently proposed swampland conjecture, the trans-Planckian censorship conjecture (TCC), also gives an upper bound to the energy scale during inflation \cite{Bedroya:2019snp, Bedroya:2019tba}. The trans-Planckian censorship conjecture postulates that any consistent theory of quantum gravity must forbid quantum fluctuations with wavelengths shorter than the Planck length from being redshifted by cosmological expansion to wavelengths where they become classical perturbations. In Ref. \cite{Lin:2019pmj}, Lin and Kinney generalize the TCC  to certain \textit{k}-inflation scenario and derive the corresponding bound on the inflationary energy scale. It is then an interesting subject to compare the energy bounds derived from different swampland conjectures in the \textit{k}-inflation scenario.

\section{conclusion}\label{Sec_Conclusion} 

In this work we examine the general embedding of a $ P(X, \varphi) $ inflationary theory into a two-field theory with curved field space. By promoting the canonical kinetic term of the field $ X $ to an auxiliary field $ \chi $ and introducing a small kinetic term to $ \chi $, the resulting two-field action is    
\begin{equation}
\label{embedding_k_inflation*}
S=\int d^{4}x\sqrt{-g}\left[\frac{M_P^2}{2}R+P(\varphi, \chi)-\frac{1}{2\Lambda^{6}}(\partial \chi)^2 +(X-\chi)P_{\chi}\right],  
\end{equation}
which reduces back to the original $ P(X, \varphi) $ theory when $ \Lambda^6 \rightarrow \infty $.  
The resulting two-field theory is consistent to a special type of two-field model with heavy entropic fluctuations, in which the heavy field can be integrated out at full action level. In the two-field model generated from this embedding, both the turning rate and effective mass of the entropic fluctuations are proportional to the single parameter $ \Lambda^6 $, which is the same parameter used in the action to turn off the kinetic term of the auxiliary field. 

However, in this type of model, the effective mass of the heavy field is bounded by the cutoff scale of the inflationary $ P(X, \varphi) $ theory, so $ \Lambda $ is bounded from above making this embedding an approximation. The bounds on $ \Lambda $ are given by
\begin{equation}
\label{bounds_on_Lambda_conclusion}
\frac{2X^2}{M_P^2 \epsilon}\frac{c_S^2}{1-c_S^2} \ll  \Lambda^6   < \frac{4\sqrt{2}\pi X}{\sqrt{P_{XX}}}\frac{c_S^{7/2}}{1-c_S^2}, 
\end{equation}
where the lower bound is from the condition that the effective mass of the heavy field has to be much greater than the Hubble parameter to be integrated out. 

We quantify the effect of a finite $ \Lambda $ on this embedding picture and derive the first potential slow-roll parameter to the next leading order in terms of quantities defined in a $ P(X, \varphi) $ theory. With $ c_S^2 \ll 1 $ and assuming small flow parameters, it is given as
\begin{equation} 
\label{final_final_epsilonV*}
\epsilon_V \simeq \epsilon \left[1 +\frac{\Lambda^6  P_{XX}}{36 H^2} \frac{1}{c_S^2} -  \frac{\beta}{3}+ \left(\frac{\beta}{6}\right)^2+  \frac{1}{9}(2\beta^2 c_S^2 -3 \beta)(\frac{3\lambda}{\Sigma}-\frac{1}{2}+\frac{\epsilon}{12c_S^2})\right], 
\end{equation}
where $ \beta \equiv \dot{\varphi}P_{\varphi X}/(HP_X) $ and $ c_S $ is the speed of sound derived from the $ P(X, \varphi) $ theory. 
We also derive the first order correction to the speed of sound given by the two-field picture. Assuming  $ c_S^2 \ll 1 $ and small flow parameters, it is given as
\begin{equation} 
\label{SpeedofSound_NL_Approx*}
c_{s(2-field)}^{-2} \simeq c_S^{-2}-\frac{6XH^2}{\Lambda^6 P_X}\left(2\beta + c_S^2 \beta^2  \right).  
\end{equation}
To have $ c_{s(2-field)}^{-2} \simeq c_S^{-2} $, we show that the deviation of $ \epsilon $ and $ \epsilon_{V} $ can only be dominated by the term proportional to $ \Lambda^6 $. From the upper bound of $  \Lambda^6  $, we then have the upper bound of $ \epsilon_{V} $ in this type of model as 
\begin{equation}
\label{Bound_on_epsilon_V*}
\epsilon_{V} < \frac{\pi}{9}\sqrt{(1-c_S^2)c_S}\epsilon^{3/2}\frac{M_P}{H}. 
\end{equation}
This result is of particular interest since for this type of model, satisfying the de Sitter conjecture gives an upper bound on the inflationary energy 
\begin{equation}
\label{Bound_on_H_UV*}
H < \frac{\pi}{9}\sqrt{(1-c_S^2)c_S}\epsilon^{3/2}M_P,    
\end{equation} 
which can be compared with another upper bound on the inflationary energy scale from the trans-Planckian censorship conjecture in \textit{k}-inflation scenario  \cite{Lin:2019pmj}.  Besides the immediate comparison of the two upper bounds from two different swampland conjectures, it is also interesting to see what kind of conditions emerge to satisfy other swampland conjectures and the other well-known bounds, for instance, the distance conjecture \cite{Ooguri:2006in} and the Lyth bound \cite{Lyth:1996im}. This general embedding offers a framework to examine the relations between those different swampland conjectures and other well-known bounds in $ P(X, \varphi) $ inflationary theory. The study of the relations between the swampland conjectures and the corresponding bounds in \textit{k}-inflation will be in the next work. 

\section*{Acknowledgements}
The authors would like to thank William H. Kinney and Dejan Stojkovic for very useful comments throughout the writing of this article.

\appendix
\section{Speed of sound to the next leading order \label{Appendix}}

In the following, we calculate the speed of sound $c_{s(2-field)}^{-2} $ in terms of quantities can be defined in the $ P(X, \varphi) $ to order $\mathcal{O}(\Lambda^{-6})$. First, by using the formulas for the speed of sound (\ref{multiF_SoundSpeed}) and the effective mass of the isocurvature perturbations (\ref{EFF_ISO_Mass}), together with the action of the embedding two-field model (\ref{embedding_k_inflation}), to order $\mathcal{O}(\Lambda^{-6})$ we have 
\begin{equation}
\label{SpeedofSound_NL_derive_in_chi}
\begin{split}
c_{s(2-field)}^{-2} = 1+&\frac{4\Omega^2}{M_{eff}^2} \\ \simeq 1+ &\frac{16\chi^2 P_{\chi \chi}^2}{P_{\chi \chi}\left[4\dot{\varphi}^2P_{\chi}+(\dot{\varphi}^4-4\chi^2)P_{\chi \chi}\right]+2\dot{\varphi}^2(\dot{\varphi}^2-2\chi)P_{\chi}P_{\chi \chi \chi}}   \\
+ &16\chi \dot{\chi}  P_{\chi \chi} \left\lbrace - P_{\chi \chi} \left[-2\dot{\varphi}^3(P_{\varphi}-\chi P_{\chi \varphi})P_{\chi \chi} + 3\chi \dot{\varphi}^2 \dot{\chi} P_{\chi \chi}^2 + 2\chi \dot{\chi}  P_{\chi \chi}(2P_{\chi}+\chi P_{\chi \chi})  \right. \right. \\
&\left. \qquad \qquad \qquad \quad - 4\dot{\varphi}(P_{\varphi}-\chi P_{\chi \varphi})(2P_{\chi}+\chi P_{\chi \chi})-8\dot{\varphi} \chi^2 P_{\chi}P_{\chi \chi \varphi} \right] \\
& \left. \qquad \qquad \quad + 2 P_{\chi}P_{\chi \chi \chi} \left[2\dot{\varphi}(-2\chi+\dot{\varphi}^2)(-P_{\varphi}+\chi P_{\chi \varphi})+ \chi \dot{\chi}  P_{\chi \chi}(-2\chi+3\dot{\varphi}^2)\right] \right\rbrace  \\
&/ \Lambda^6 P_{\chi} \left\lbrace P_{\chi \chi} [4\dot{\varphi}^2 P_{\chi}+ (-4\chi^2+\dot{\varphi}^4 )P_{\chi \chi}]-2\dot{\varphi}^2(-2\chi+\dot{\varphi}^2)P_{\chi}P_{\chi \chi \chi}\right\rbrace^2.
\end{split}
\end{equation}
Notice that Eq. (\ref{VN_turning}) is used to evaluate the turning rate $ \Omega $ to have the above expression, which is the reason why the  $\mathcal{O}(\Lambda^0) $ terms is different from the form in Eq. (\ref{SoundSpeed_embedding}). If we substitute $ \chi \simeq X $, the $\mathcal{O}(\Lambda^0) $ terms give the speed of sound formula of $ P(X, \varphi) $ theory (\ref{SpeedofSound*}) as it should be. 

Here, to have the next order result, we substitute the constraint with $\mathcal{O}(\Lambda^{-6})$ correction (\ref{chi=X_approx}) and keep only terms to $\mathcal{O}(\Lambda^{-6})$. After some algebra we have 
\begin{equation}
\label{SpeedofSound_NL_derive_in_X}
\begin{split}
c_{s(2-field)}^{-2}  \simeq& 1+ \frac{2X P_{XX}}{P_X} -\frac{4\Box X}{\Lambda^6 P_X}   \\
&+ \frac{\dot{X}}{\Lambda^6 \dot{\varphi}P_X^2} \left[4P_{\varphi}(1+\frac{XP_{XX}}{P_X})-\dot{\varphi}(\dot{\varphi} P_{X\varphi}+\dot{P_X})\right] \\
& + \frac{\dot{X}}{\Lambda^6 P_X^3} \left[-2X P_{XX}\dot{P_X}+2X P_{X}\dot{P_{XX}}\right]. 
\end{split}
\end{equation}
To rewrite it into flow parameters, we first use Eq. (\ref{EoM_non_can_in_parameters}) to rewrite $ \dot{X} $ as
\begin{equation}
\label{X_dot}
\dot{X}=\dot{\varphi}\ddot{\varphi}=\left(\frac{\eta}{2}-\epsilon-\frac{\tilde{\kappa}}{2}\right)H \dot{\varphi}^2 \equiv 2HXA,  
\end{equation}
where $ A \equiv \left(\frac{\eta}{2}-\epsilon-\frac{\tilde{\kappa}}{2}\right)  $ for simplification. And from Eq. (\ref{X_dot}), we have 
\begin{equation}
\label{X_doubleDot}
\ddot{X} =2HX\left[\dot{A}+HA(2A-\epsilon)\right], 
\end{equation}
where by using the flow parameters $ \kappa \equiv \dot{\eta}/(H\eta)$ and $ \tilde{\lambda} \equiv \dot{\tilde{\kappa}}/(H\tilde{\kappa}) $, $ \dot{A} $ is given by
\begin{equation}
\label{A_dot}
\dot{A}=H\left(\frac{\kappa \eta}{2} -\epsilon \eta -\frac{\tilde{\kappa}\tilde{\lambda}}{2}\right).  
\end{equation}
Using Eqs. (\ref{X_dot}, \ref{X_doubleDot}) we have
\begin{equation}
\label{Box_X}
\Box X=- \ddot{X} -3H\dot{X}= -2H^2X \left[(\frac{\kappa \eta}{2}-\epsilon \eta -\frac{\tilde{\kappa} \tilde{\lambda}}{2})+(\frac{ \eta}{2}-\epsilon -\frac{\tilde{\kappa}}{2})(3+ \eta-3\epsilon -\tilde{\kappa})\right] \equiv -2H^2X \Delta. 
\end{equation}
Next, to rewrite $ \dot{P_{XX}} $, we take the time derivative of the relation
$ 2X P_{XX}= P_{X}(c_S^{-2}-1) $ to have 
\begin{equation}
\label{P_XX_dot_relation}
 2X \dot{P_{XX}}= \tilde{\kappa} P_X H (c_S^{-2}-1) - 2sH P_X c_S^{-2} -4AHXP_{XX}, 
\end{equation}
where $ s \equiv \dot{c_S}/(Hc_S) $ is the first speed of sound flow parameter. Using the relations above and in Section \ref{Sec. background} A, we have $ c_{S(2-field)}^{-2} $ to order $ \mathcal{O}(\Lambda^{-6}) $ in terms of quantities defined in the $ P(X, \varphi) $ theory as
\begin{equation}
\label{SpeedofSound_NL_in_A}
\begin{split}
c_{S(2-field)}^{-2} \simeq \: c_S^{-2} + & \frac{8H^2 X}{\Lambda^6 P_X}\Delta  \\
+& \frac{2H^2 X}{\Lambda^6 P_X} A\left[2\left(3+A+\tilde{\kappa}\right)\left(\frac{1+c_S^2}{c_S^2}\right)-\beta-\tilde{\kappa}\right] \\
 -& \frac{4H^2 X}{\Lambda^6 P_X} A\left[A\left(\frac{1-c_S^2}{c_S^2}\right)+\frac{s}{c_S^2}\right]. 
\end{split}
\end{equation}
Notice by using Eq. (\ref{new_slow_roll_para}), we can replace $ \tilde{\kappa} $ and also rewrite $ A $ as  
\begin{equation}
\label{A_in_beta}
A=\frac{c_S^2}{1+c_S^2}(\eta-2\epsilon-\beta), 
\end{equation}
which is a more convenient form for the common scenario that all the flow parameters are small except $ \beta $ since it is technically not a flow parameter. With this approximation, $ \epsilon, \eta, \kappa, s, \tilde{\lambda} \ll 1 $, Eq. (\ref{SpeedofSound_NL_in_A}) reduces to
\begin{equation}
\label{SpeedofSound_NL_in_A_2}
c_{S(2-field)}^{-2} \simeq c_S^{-2}-\frac{XH^2}{\Lambda^6 P_X}\frac{2\beta\left[6(1+3c_S^2)(1+c_S^2)+\beta c_S^2 (3-11c_S^2)\right]}{(1+c_S^2)^2}.
\end{equation}
If we further assume $ c_S^2 \ll 1  $ we have Eq. (\ref{SpeedofSound_NL_Approx}) used in Sec. \ref{Sec_de_Sitter}
\begin{equation} 
\label{SpeedofSound_NL_ApproxAppendix}
c_{S(2-field)}^{-2} \simeq c_S^{-2}-\frac{6XH^2}{\Lambda^6 P_X}\left(2\beta + c_S^2 \beta^2  \right).  
\end{equation}

\bibliographystyle{apsrev4-1}
\bibliography{general_embedding}

\end{document}